# Hydrogen-induced hardening of a high-manganese twinning induced plasticity steel

Heena Khanchandani[1*], Dirk Ponge[1], Stefan Zaefferer[1], Baptiste Gault[1,2*]

[1]Department of Microstructure Physics and Alloy Design, Max-Planck-Institut für Eisenforschung, Max-Planck-Str. 1, 40237 Düsseldorf, Germany.

[2]Department of Materials, Royal School of Mines, Imperial College, Prince Consort Road, London SW7 2BP, United Kingdom.

*Corresponding authors: h.khanchandani@mpie.de & b.gault@mpie.de

**Abstract:** High-manganese twinning-induced plasticity (TWIP) steels exhibit high strain hardening, high tensile strength, and high ductility, which make them attractive for structural applications. At low tensile strain rates, TWIP steels are prone to hydrogen embrittlement (HE). Here though, we study the hardening and strengthening resulting from electrochemical hydrogen-charging of a surface layer of a Fe-26.9Mn-0.28C (wt.%) TWIP steel. We observed a 20% increase in yield strength following the electrochemical hydrogen-charging, accompanied by a reduction in ductility from 75% to 10% at a tensile strain rate of $10^{-3}s^{-1}$. The microstructural evolution during tensile deformation was examined at strain levels of 3%, 5% and 7% by electron backscatter diffraction (EBSD) and electron channeling contrast imaging (ECCI) to study the dislocation structure of the hardened region. As expected, the microstructure of the hydrogen-hardened and the uncharged regions of the material evolve differently. The uncharged areas show entangled dislocation structures, indicating slip from a limited number of potentially coplanar slip systems. In contrast, hydrogen segregated to the grain boundaries, revealed by the deuterium-labelled atom probe tomography, delays the dislocation nucleation by blocking dislocation sources at the grain boundaries. The charged areas hence first show the formation of cells, indicating dislocation entanglement from more non-coplanar slip systems. With increasing strain, these cells dissolve, and stacking faults and strain-induced ε-martensite are formed, promoted by the presence of hydrogen. The influence of hydrogen on dislocation structures and the overall deformation mechanism is discussed in details.

# 1 Introduction

TWIP steels are a class of high manganese (typically over 18 wt.%), austenitic steels, i.e. with a face centered cubic (FCC) crystal structure [1,2]. They deform by dislocation glide along with the continuous formation of mechanical twins during deformation [3]. The increase in twin density leads to a very high strain hardening rate, which is often referred to as the 'dynamic Hall-Petch effect' [4]. TWIP steels hence have an excellent combination of high tensile strength and ductility, making them potentially suitable for a wide range of applications in the automotive industry, line pipe production and ship building for instance [5].

However, TWIP steels are highly susceptible to HE, as often demonstrated by hydrogen-induced delayed fracture in deep-drawn cups [2,6,7]. The degradation in mechanical properties due to hydrogen has been studied in an array of TWIP or TRIP (transformation induced plasticity) steel compositions including Fe-18Mn-0.6C (wt.%) [8], Fe-18Mn-1.2C (wt.%) [9], Fe-23Mn-0.5C (wt.%) [10] and Fe-22Mn-0.6C (wt.%) [11]. The plausible mechanisms proposed in these studies responsible for HE in TWIP steels include intergranular cracking associated to the promotion of deformation twinning [8–10] and martensitic transformation [8] induced by hydrogen.

Despite its deleterious effects, hydrogen is an interstitial element in austenite and has also been reported to cause solid solution strengthening [12,13]. However, there are only a limited number of studies focusing on the strengthening effect of hydrogen. It is the interaction of hydrogen with structural defects formed during the deformation that decides if the material is embrittled or strengthened by the presence of hydrogen [9,14], which highly depends on the strain rate. These studies require the precise measurement and characterization of hydrogen, which is challenging, leaving numerous questions opened [15–18]. Therefore, the resultant microstructural changes associated to the presence of hydrogen must be examined thoroughly.

Here, we examined the strengthening effect of hydrogen in a model Fe-26.9Mn-0.28C (wt.%) TWIP steel. Hydrogen was introduced into the samples by cathodic hydrogen charging. We performed tensile tests at a strain rate at which the effect of HE is not expected to be prominent. Subsequently, we investigated the structural defects formed during the tensile deformation with and without hydrogen charging. Since the hydrogen-charged sample exhibited a 20% increase in the yield strength with the total elongation to fracture of 10%, we chose three lower tensile strain levels, 3%, 5% and 7% in order to study the dislocation structure of the region hardened by hydrogen. The charging with hydrogen was confirmed quantitatively by thermal desorption spectroscopy (see supplementary information, Figure S1, for details). We mapped microstructural changes that result from the tensile deformation of the hydrogen-charged and the uncharged samples by using electron channeling contrast imaging (ECCI) in the scanning-electron microscope (SEM), enabling the precise identification of dislocation structures across wide areas [19]. Atom probe tomography (APT) was employed to study the hydrogen distribution at or near structural defects such as grain boundaries, dislocations and stacking faults. APT provides in principle the spatial distribution of specific chemical species with very high chemical sensitivity up-to sub-nanometer resolution [20] and has been used to study hydrogen present at specific microstructural features [21–24]. This correlative approach helps to identify the defect structure and composition, thereby facilitating to understand the mechanisms involved in strengthening by hydrogen in TWIP steels. The current study further shows that TWIP steels develop a harder surface layer similar to a coating upon hydrogen ingress.

## 2 Materials and Methods

A Fe-26.9Mn-0.28C (wt.%) TWIP steel was produced by strip casting [25]. Homogenization annealing was carried out at 1150 °C for 2 hours [25]. Following the removal of surface oxides and of regions possibly decarburized or demanganized, it was cold-rolled to achieve a 50% thickness reduction, and subjected to recrystallization annealing at 800 °C for 20 minutes, followed by water cooling to room temperature. The samples were cut into dog-bone tensile sample geometry with gauge dimensions of $4 \times 2 \times 1$ mm$^3$ by using electrical discharge machining (EDM). Next, they were mechanically ground by #600 and #1000 grinding paper, followed by mechanical polishing with 3 µm diamond paste for 25 minutes. The final sample thickness is approx. 900 µm. Subsequently, a mechano-chemical polishing step was carried out with 50 nm colloidal silica suspension for 40 minutes. The samples were finally cleaned with ethanol and dried with compressed air. All surfaces of the samples were thoroughly cleaned to remove corrosion prior to hydrogen charging. The side surfaces were hence finished with #400 grinding paper, while the top and bottom surfaces were prepared as described above, until the mechano-chemical polishing step, in order to facilitate the hydrogen ingress into the sample during subsequent cathodic hydrogen charging [26].

The cathodic hydrogen charging was conducted in an electrolytic cell in which an aqueous solution of 0.05 M $H_2SO_4$ was used as electrolyte [27–29], and 1.4 g/l of thiourea ($CH_4N_2S$) was added as hydrogen recombination inhibitor, thereby enhancing the hydrogen ingress [30]. The charging solution was magnetically stirred at a speed of 10 rotations per minute throughout the charging process in order to maintain a constant rate of charging reaction. A DC voltage of 1.5V was applied to the cell where the sample served as cathode and a Pt wire as anode [21,31]. The cathodic current density during charging was 75 A/m$^2$. The cathodic hydrogen charging set-up is shown schematically in Figure 1(a).

The hydrogen charging was carried out for 5 days at ambient temperature conditions. The sample surface was then cleaned by a 5 minutes polishing using 50 nm colloidal silica suspension and then by rinsing with ethanol. This process removed approximately 200 nm of material. Following this, the sample showed a clean, mirror-like metallic surface. The tensile tests were hence conducted within 10 minutes after the charging had finished at a constant strain rate of $10^{-3}$ s$^{-1}$ at room temperature using a Deformation Devices System DDS-3 (Kammrath & Weiss GmbH). The ARAMIS v6.3.0 software was used for the data extraction and analysis of tensile tests.

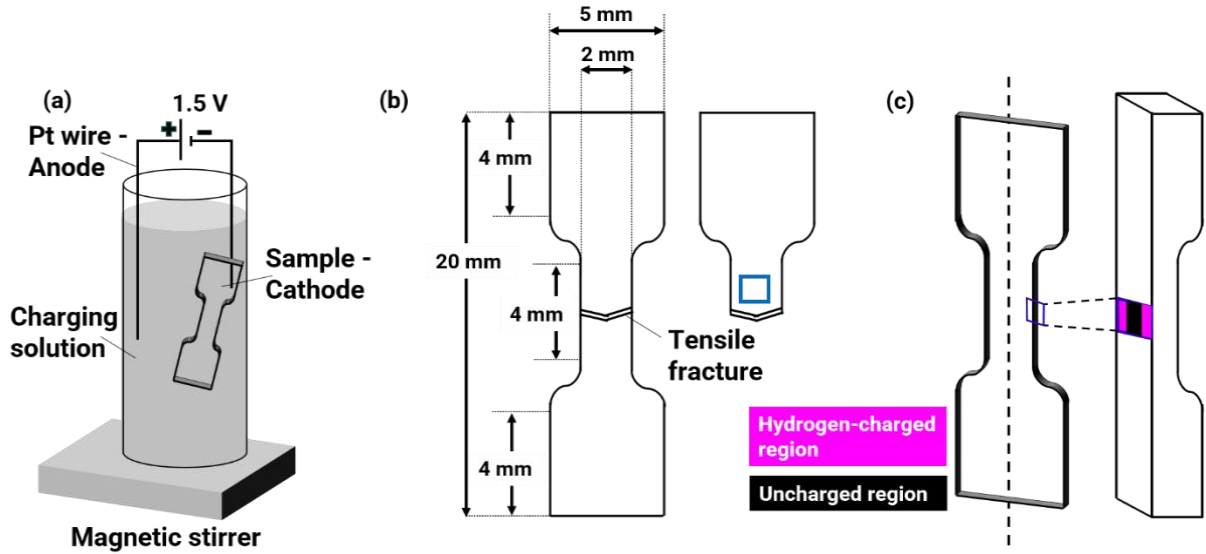

*Figure 1. (a) A schematic of cathodic hydrogen charging set up; (b) a schematic depicting a tensile sample deformed to fracture; (c) a schematic of tensile sample strained to an intermediate level of tensile strain followed by a vertical cut to examine the cross section containing the hydrogen-charged region, that is within ~100 µm close to the side edges of the sample surface (indicated in pink) and the uncharged region (indicated in black).*

The fracture surfaces of tensile deformed samples were examined using a Zeiss Sigma 500 SEM at a working distance of 10mm, an accelerating voltage of 15kV and a beam current of 9nA. ECCI studies were carried out using a Zeiss Merlin SEM, equipped with a retractable 4Q backscattered electrons detector. The microscope was operated at an accelerating voltage of 30kV, a beam current of 2nA and a working distance of 7mm [19]. The dislocation structures of the tensile sample deformed to fracture were examined from the region delineated by the blue box in Figure 1(b), approx. 500 µm away from the fracture surface.

Three hydrogen-charged samples were deformed to tensile strains of 3%, 5% and 7%, respectively at a strain rate of $10^{-3}$ $s^{-1}$. Since the effect of HE is more pronounced at $10^{-4}$ $s^{-1}$ or lower strain rates in TWIP steels [10], a high tensile strain rate of $10^{-3}$ $s^{-1}$ was selected to avoid the diffusion of hydrogen by the moving dislocations during deformation and hence to study the strengthening effect of hydrogen. This also prevents the diffusion of hydrogen towards the center region of the sample during deformation. This region therefore remains uncharged and can be studied as such. The deformed samples were immediately vertically cut into two halves by using EDM, as shown schematically in Figure 1(c). The obtained cross sections were mounted in electrically conductive resin at 180°C for 6 minutes. The mounting of the hydrogen-charged samples was performed following the tensile deformation, hence this exposure to higher temperature does not influence the interaction of hydrogen with dislocations that formed the dislocation structure during deformation which is the main focus of the current study. Subsequently, they were polished until the final mechano-chemical polishing step on the same day and their dislocation structures were examined by ECCI within 12 hours of the tensile test. A homogeneous distribution of hydrogen could not be achieved even after 5 days of charging. We hence chose the region highlighted in pink in Figure 1(c) within ~100 µm close to the side edges of the sample surface to examine the dislocation structure obtained by deformation in the presence of hydrogen. This region is referred to as the hydrogen-charged region in the current study. The center region highlighted in black across the cross-section, as shown in Figure 1(c) was chosen to examine the dislocation structure formed during deformation without hydrogen and is referred to as the uncharged region. Approximately 20 grains were randomly chosen from each of the two regions for examining the dislocation structures statistically. Four stage tilt angles were selected, i.e., ~ 0°, ~ -1°, ~ 5° and ~ 10° to examine the grains with different orientations and at least 5 grains were examined which were in two-beam diffraction conditions at each of the tilt angles. In the two-beam diffraction condition, only one set of lattice planes is tilted into the diffraction condition [19]. These conditions allow clear observation of dislocations or stacking faults as bright features on a dark background.

In order to obtain the orientations and morphologies of the grains, EBSD orientation mapping was performed using a Zeiss Sigma 500 SEM equipped with an EDAX/TSL system with a Hikari camera at an accelerating voltage of 15kV, a beam current of 9nA, a scan step size of 0.5µm, a specimen tilt angle of 70°, and a working distance of 14mm [32]. EBSD data analyses were performed by using TSL OIM Analysis 7.0.

APT specimens were prepared from targeted grains within the hydrogen-charged region of deformed samples, selected by ECCI, by using a FEI Helios NanoLab 600i dual-beam FIB/SEM using the approach for site-specific lift-out outlined in Ref. [33]. Analysis by APT of hydrogen in deep traps inside Fe-based materials had been reported previously [34,35], despite sample storage and preparation at room temperature. APT specimens were further charged with deuterium to help reveal the distribution of hydrogen at different microstructural features, particularly if diffusible hydrogen had already been released. Deuterium was used to avoid overlap with the hydrogen signal originating from the ionization of residual gases from the ultrahigh-vacuum chamber [31,35,36]. Deuterium charging was performed in a gas-charging chamber described in ref. [37]. APT specimens were exposed to 250 mbar of deuterium gas at room temperature for 6 hours, subsequently quenched to cryogenic temperatures in a precooled ultra-high vacuum suitcase [38] and transferred to the LEAP 5000XS or XR instrument (CAMECA Instruments Inc. Madison, WI, USA) for analysis. APT analysis was performed in voltage pulsing mode at a set point temperature of 70K, 0.4% detection rate, 15% pulse fraction and 100 kHz pulse repetition rate, using conditions that were shown to maximize the reliability of the results as discussed in Ref. [31].

## 3 Results:

*3.1 Initial microstructure*

The initial microstructure of the recrystallized, uncharged Fe-26.9Mn-0.28C (wt.%) TWIP steel is displayed in Figure 2(a) by an EBSD-measured inverse pole figure (IPF) map with respect to the normal direction (ND). 95% of all grain boundaries are high angle grain boundaries, marked in red, and 36.6% of these are Σ3 twin boundaries, marked in black. The material is fully austenitic and its grain size distribution is depicted in Figure 2(b).

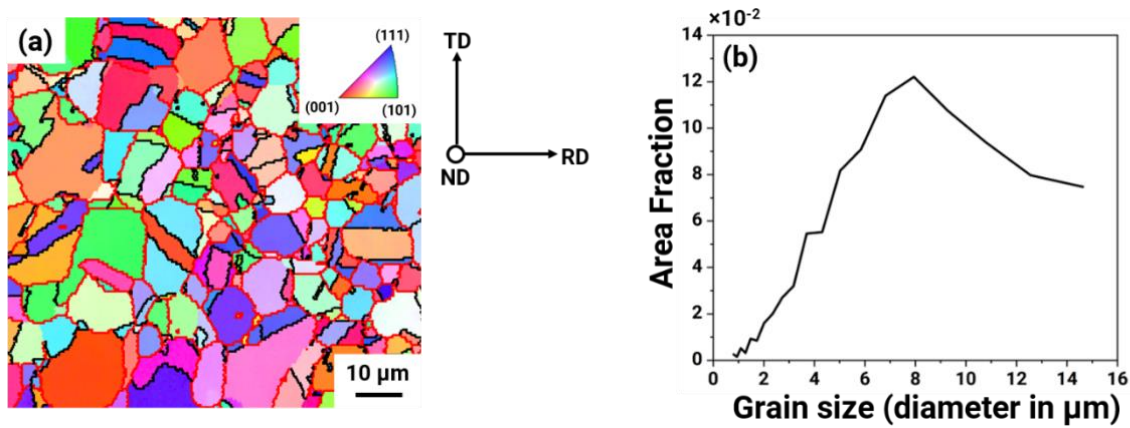

*Figure 2. (a) EBSD – ND inverse pole figure (IPF) map of the recrystallized Fe-26.9Mn-0.28C (wt.%) TWIP steel. TD – transverse direction; ND – normal direction out of the plane; RD – rolling direction; (b) the grain size distribution.*

We performed energy dispersive X-ray spectroscopy (EDX) mapping, shown in the supplementary information (Figure S2), to confirm the homogeneous distribution of manganese across the sample, from the near surface to the center region. The composition profile depicted in Figure S2(h) indicates an approximately constant manganese content of ~26 wt.% throughout.

*3.2 Tensile test*

Three hydrogen-charged and three uncharged samples were subjected to tensile tests. One of the engineering stress-strain curves is shown in Figure 3. The other two curves from each of the hydrogen-charged and the uncharged samples are provided in the supplementary information (Figure S3). They show very closely similar behavior. The yield strength of the uncharged sample was 250 MPa while it was 300 MPa for the hydrogen-charged sample, as highlighted in the close-up in the inset of Figure 3, the yield strength was hence increased by 20% after hydrogen charging. The hydrogen-charged sample also shows a small plateau at the yield point following the yielding. The total elongation to fracture of the uncharged sample was ~75% (black curve) which was reduced to ~10% (pink curve) after hydrogen charging.

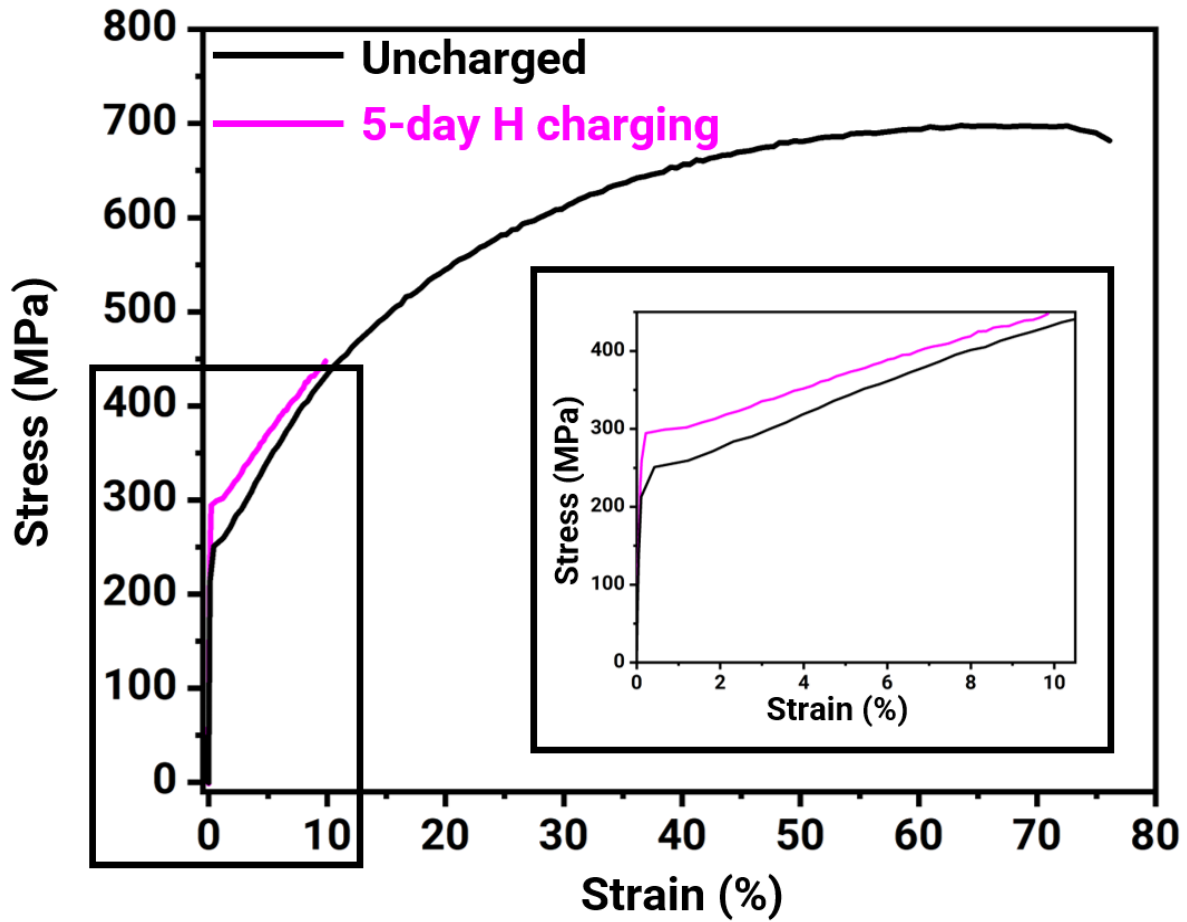

*Figure 3. The engineering stress-strain curve of the uncharged (in black) and the hydrogen-charged (in pink) samples (the inset shows the enlarged region delineated by black box).*

### 3.4 Fractography

The fracture surface of the uncharged sample consists of dimples, indicating ductile behavior, as shown by SEM imaging in Figure 4(a). Figure 4(b) shows the fracture surface of the hydrogen-charged sample where the outermost approx. 30 µm region, close to the side edges, shows evidence of intergranular fracture, as revealed by the inset in Figure 4(b). The fracture mode in the center of the hydrogen-charged sample was dominated by dimples associated to ductile fracture.

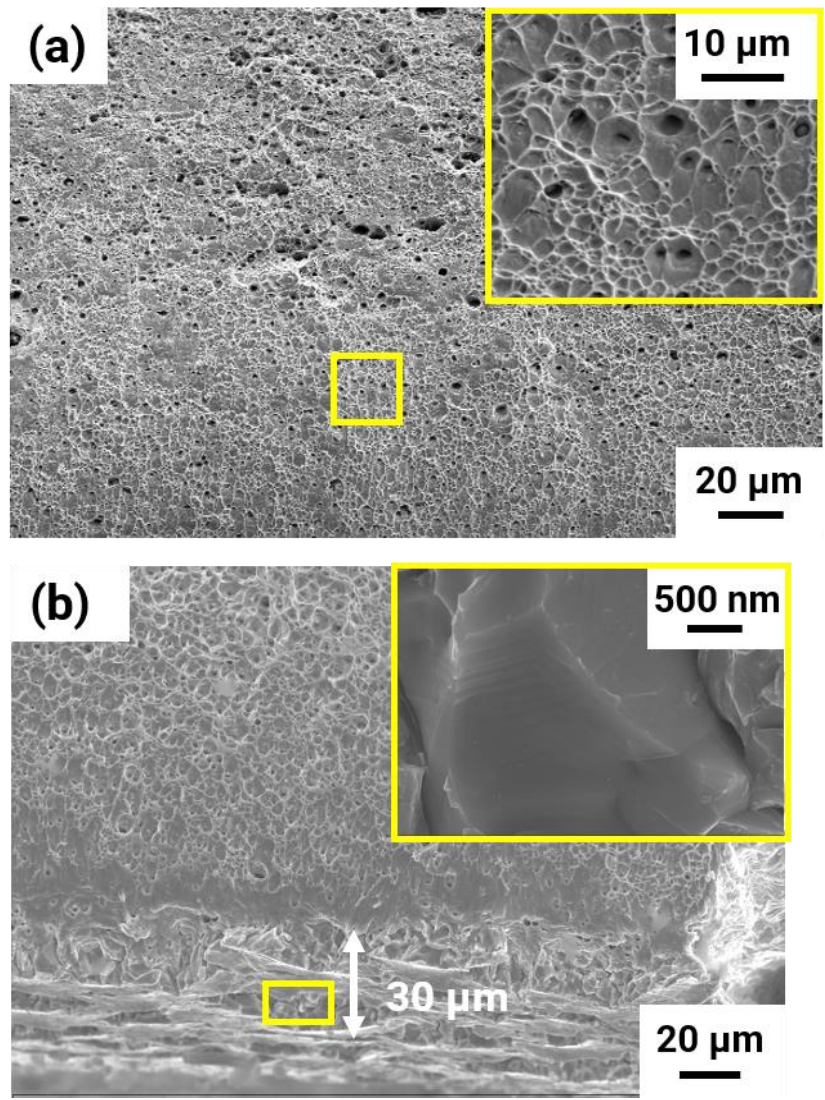

*Figure 4. (a) SEM image of the fracture surface of the uncharged sample strained to 75 % of tensile strain, where the inset shows the enlarged region delineated by yellow box; (b) SEM image of the fracture surface of the hydrogen-charged sample strained to 10 % of tensile strain, where the inset shows the enlarged region delineated by yellow box. The original sample surface through which hydrogen diffused into the sample is at the bottom edge of the image.*

The brittle fracture due to hydrogen charging was hence limited to the region closest to the sample's surface exposed to the hydrogen charging, which can be explained by the slow diffusivity of hydrogen within austenite, pointed out in ref. [39–41] for instance. The microstructures of these samples deformed to tensile fracture at a distance of approx. 500 μm from the fracture surface are shown in the supplementary information, Figure S4.

### 3.5 Electron-channeling contrast imaging of the deformed samples
#### 3.5.1 3% tensile strained sample

Figure 5(a) shows the ECC micrograph of a grain from the hydrogen-charged region, at a distance of 20 μm from the surface of the sample, deformed to 3% tensile strain. The micrograph evidences dislocations forming cell-like structures. In contrast, the ECC micrograph of a grain from the uncharged region of the same sample shows entangled dislocations, as shown in Figure 5(b).

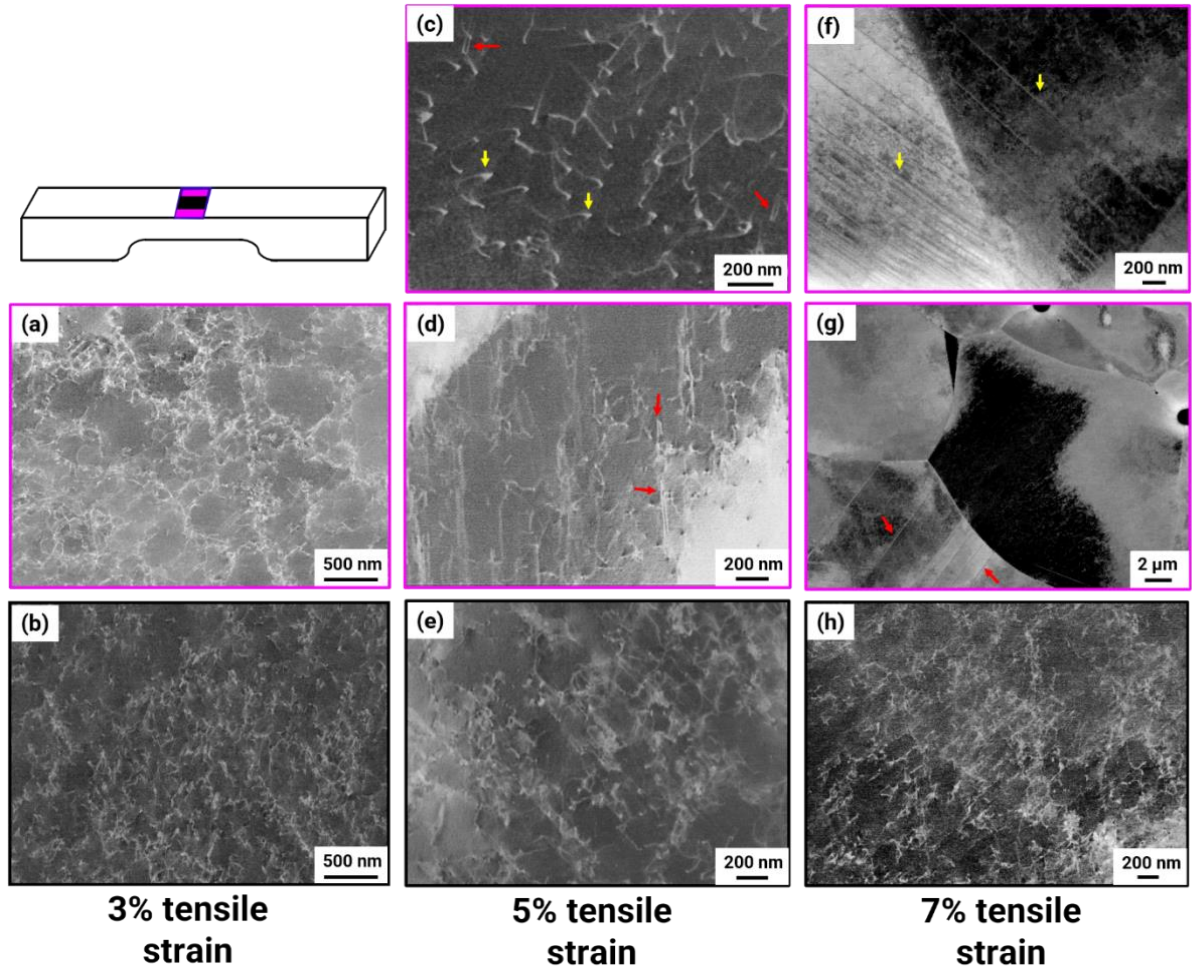

*Figure 5. ECC micrographs of deformation microstructures from different areas and deformation degrees. Pink frames indicate images from the hydrogen-charged region, black from the uncharged region. Left to right columns are from 3, 5 and 7% tensile strained samples. (a) dislocation structures with cell morphology; (b) entangled dislocations; (c-d) dislocation partials indicated by yellow arrows and dislocation loop dipoles indicated by red arrows; (e) entangled dislocations; (f-g) long, extended stacking faults indicated by yellow arrows and the nucleation of features interpreted as ε-martensite phase indicated by red arrows; (h) entangled dislocations.*

### 3.5.2  5% tensile strained sample

Figures 5(c-d) show ECC micrographs of two different grains from the hydrogen-charged region, at a distance of 15 µm from the surface of the sample, deformed to 5% tensile strain. The features indicated by yellow arrows in Figure 5(c) are two closely-spaced dislocation partials bounding a stacking fault, indicating planar slip. The features indicated by red arrows in Figures 5(c-d) are likely dislocation loop dipoles, as there is no stacking fault visible between them. Such features are typically formed due to double cross slip. An ECC micrograph of a grain from the uncharged, center region of the same sample in Figure 5(e), shows entangled dislocations.

### 3.5.3  7% tensile strained sample

The ECC micrographs in Figures 5(f-g) are from two different grains in the hydrogen-charged region, at a distance of 10 µm from the surface of the sample, deformed to 7% tensile strain. The features pointed by yellow arrows in Figure 5(f) indicate the formation of long, extended stacking faults, potentially along with ε-martensite. The stacking faults can be distinguished from well-developed ε-martensite plates by the contrast difference in their ECC micrographs. Stacking faults are only one-atomic layer thick. Thus, in ECCI, they appear as features that are marked by an intense bright line, where the fault intersects the sample surface followed by fading contrast to one side of the line where the stacking fault dives into the material [19]. In contrast, an ε-martensite plate or a twin has an appreciable thickness and thus appears as a continuous bright band of constant brightness [19]. This is actually the case for the features marked by red arrows in Figure 5(g). Based on the previous work on a similar alloy by An *et al*. [42], we interpret these features as a nucleus of ε-martensite. As suggested by the literature, the formation mechanisms of both ε–martensite and twins and their evolution are very similar, even if their influence on the deformation process can differ [43,44]. It must be noted that these ε-martensite plates are very thin and hence cannot be resolved by EBSD. The ECC micrograph of a grain from the uncharged region of the same sample, Figure 5(h), shows entangled dislocations and no ε-martensite phase formation.

*3.5.4 Statistical analysis*
A statistical analysis of the dislocation structures was performed by examining at least 20 randomly-selected grains from both the hydrogen-charged and the uncharged regions of each of the three samples, i.e., 3%, 5% and 7% tensile strained. The orientations of those grains were determined by using EBSD. All investigated grains show similar deformation structures, although they are from a broad range of Taylor factors, Figure S5. The inverse pole figures of the tensile direction were also drawn. No preferred orientation of grains is observed, and the texture also has no influence on the dislocation structures, as indicated by the corresponding inverse pole figures for each of the examined grains included in the supplementary text (Figure S6). At the low strains and high tensile strain rate chosen herein to study the hardening effect of hydrogen, the Taylor factor may not yet play an important role, conversely to report in e.g. Ref. [42], and the deformation may rather be governed by the Schmid factor or by the activity of hydrogen.

*3.6 Atom probe tomography*
Specimens for APT were prepared from samples that had been first electrolytically charged with hydrogen and then tensile deformed. After preparation, APT specimens were nonetheless further charged with deuterium in a gas-charging chamber, before APT analysis. Grain boundaries were first targeted, because they are loci of dislocation nucleation [45,46], including in hydrogen-charged materials [18], and intergranular fracture was observed in the hydrogen-charged region. Deuterium segregation at a grain boundary in the 3% tensile strained sample is evidenced in Figure 6. Figure 6(a) shows an ECC micrograph of the region which was lifted-out for preparing the APT specimens. Figure 6(b) displays the 3-D elemental map of the APT specimen with the grain boundary. The composition profile calculated with 0.5 nm bin width across the grain boundary is plotted in Figure 6(c), with deuterium segregation of up to $1.4 \pm 0.2$ at.% at the grain boundary, that shows also a strong manganese depletion at the boundary (approx. $- 6.5 \pm 0.5$ at.%). The overall lower manganese content in the APT measurement can be associated to the loss of the lowest evaporating field species, i.e. in this instance manganese [47,48]. We also observe an unexpected strong oxygen enrichment of up to $13.4 \pm 0.5$ at.% at the grain boundary which could have originated from the cathodic charging of the bulk specimen or from the preparation and charging of the APT specimen and during its transfers to the atom probe.

Deuterium segregation to the grain boundary confirms that the gas charging conditions allowed for some deuterium ingress into the material. Similar analyses were performed on other grains containing dislocations with a cell morphology, farther from the grain boundary, shown in Figure 6(d). The deuterium distribution is homogeneous in this APT specimen with no visible or statistically significant segregation. A binomial frequency distribution analysis was performed to assess the distribution inside the grain [48–50]. The observed frequency distributions for manganese, carbon, hydrogen and deuterium atoms in Figure 6(e) match closely the binomial distribution expected for a random distribution of solutes. The Pearson coefficient µ is used for interpreting the segregation of solutes which takes value from 0 that indicates a random distribution, i.e. no segregation and 1 that indicates a strong deviation from randomness [51]. Here, the value of µ for all the elemental species lie well below 0.1, which suggests that their distribution throughout the APT dataset was random, i.e. with no indication of segregation to microstructural defects.

Since the samples had been hydrogen-charged prior to tensile deformation, trapped hydrogen at defects might prevent efficient ingress of deuterium during the second gas loading, as suggested in ref. [35]. A frequency distribution analysis for hydrogen (i.e. excluding deuterium) similarly reveals no detectable segregation (µ~0.04-0.06) or H-composition fluctuations on the scale of 10–100 nm that could be related to the presence of e.g. cells.

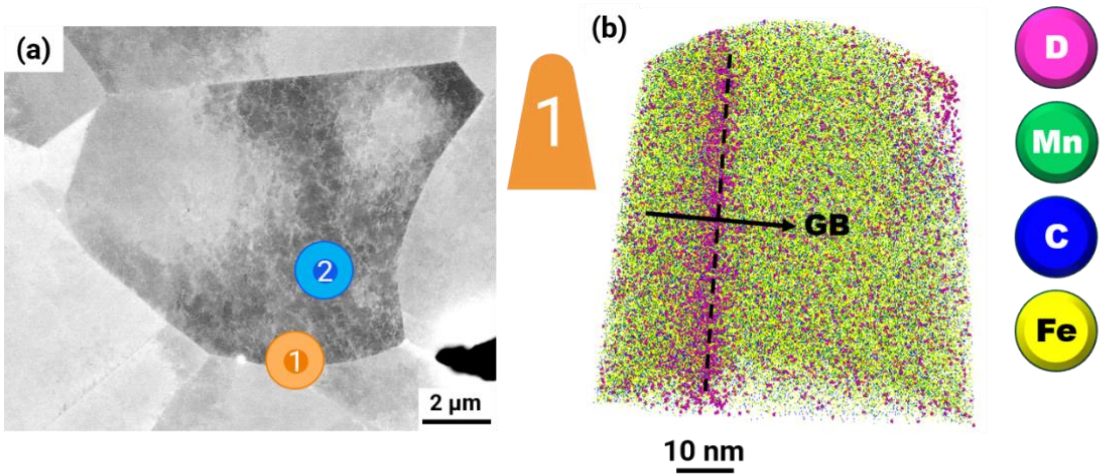
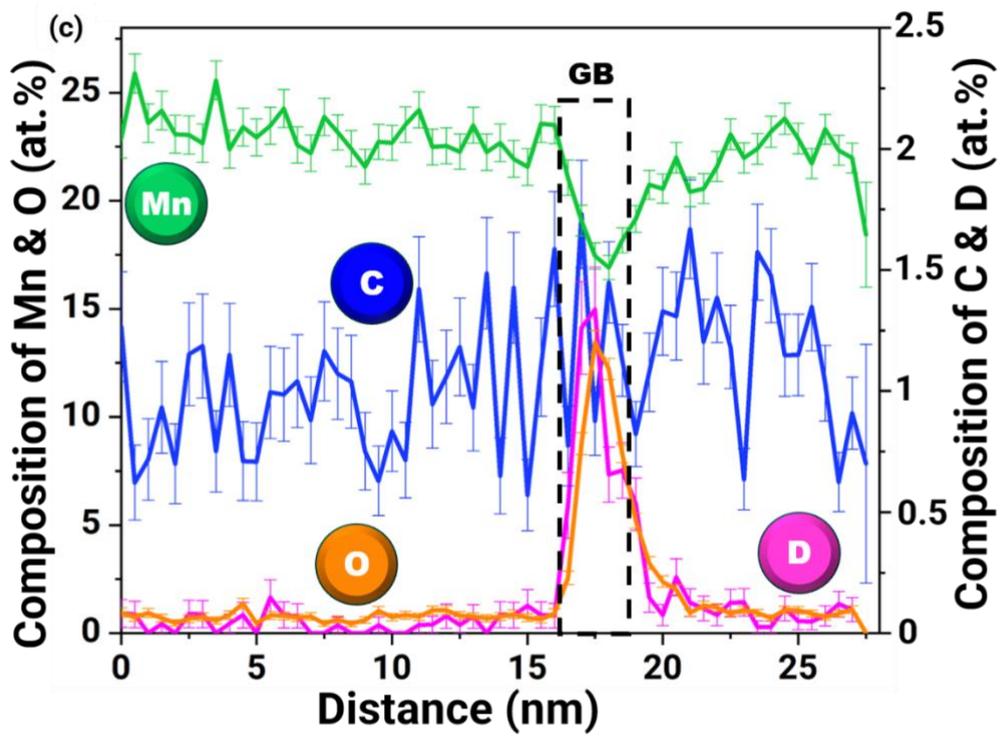
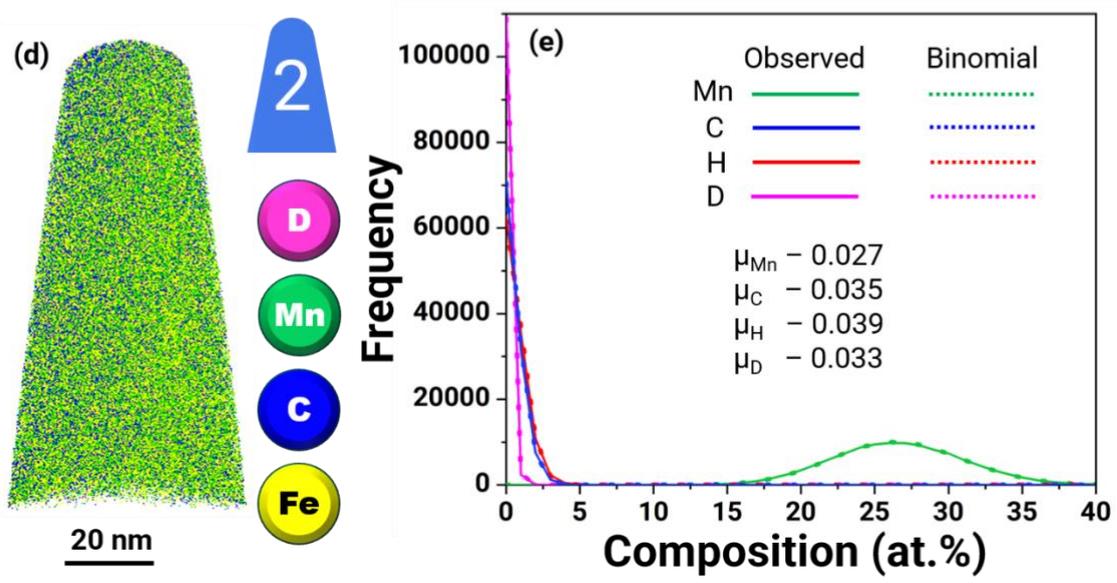

*Figure 6. (a) ECC micrograph marking the region in the hydrogen-charged region of the 3% tensile strained sample from where the APT specimens were prepared; (b) 3-D elemental map of the APT specimen showing deuterium enrichment at a grain boundary (GB); (c) the composition profile across the GB with 0.5 nm bin width, the left axis indicates composition of manganese and oxygen, the right axis that of carbon and deuterium; (d) 3-D elemental map of APT specimen prepared from the grain with dislocations exhibiting the cell morphology; (e) the binomial frequency distribution analysis corresponding to the APT dataset shown in (d).*

APT analyses were performed in the hydrogen-charged region of the 7% tensile strained sample containing stacking faults, and similarly charged with deuterium, supplementary Figure S7. No elemental segregation at dislocations or stacking faults was observed in any of these datasets which is expected at such high tensile strain rate.

## 4 Discussion

At each of the three tensile strain levels, i.e., 3%, 5% and 7%, entangled dislocations were observed in the uncharged region. This is in stark contrast with the hydrogen-charged regions: (i) dislocation cells were formed at 3%; (ii) dislocation partials and loop dipoles were formed at 5%; and (iii) long-extended stacking faults with ε-martensite were already formed at 7%. We confirmed that this microstructural evolution was due to the presence of hydrogen, and not associated to the texture or the Taylor factor. The associated mechanisms will now be discussed.

*4.1 Early stage of the deformation*

The onset of plastic deformation occurs at the yield point with the nucleation of dislocations. The yield strength of the uncharged sample was 250 MPa, whereas it was 300 MPa for the hydrogen-charged sample. A 20% increase in yield strength can be ascribed to the solid solution strengthening effect by hydrogen which is an interstitial solute element in austenite. We have to take into account that the thickness of the sample after polishing and hydrogen-charging is below 1 mm (approx. 800 µm) in the current study and only approximately 15–20% of the sample's cross section shows brittle and intergranular fracture features due to hydrogen, as shown in Figure 4(b). The side edges of the sample are hence made harder by the presence of hydrogen compared to the center, uncharged region, hence developing a hard coating on the material. To a first approximation based on the area fraction, using the rule of mixtures [52] for which 250 MPa can be used as the yield strength of the 80% uncharged region in the hydrogen-charged sample whose overall yield strength is 300 MPa, the yield strength of the 20% hydrogen-charged region is calculated as 500 MPa. Hence, a higher stress is required to deform the side edges. The higher yield strength of the hydrogen-charged region can be rationalized by the following effect. Hydrogen segregated at grain boundaries, as revealed by APT, delays the emission of dislocations by blocking the dislocation sources at or in grain boundaries, which agrees with the literature [53,54]. A higher critical stress is required to activate the grain boundary dislocation sources. Once activation (or nucleation) has been achieved, dislocations are formed and glide easily which leads to strain softening due to an increase in the mobile dislocation density and is connected with the yield point phenomena.

*4.2 3% deformation*

In the hydrogen-charged region, dislocations entangle and cells are formed, which is unexpected for materials with a low stacking fault energy, i.e. in which cross slip is unlikely. Nevertheless, cells had been observed in other TWIP steels without hydrogen charging, e.g. Fe-22Mn-0.6C (wt.%) [55] despite its relatively low stacking fault energy of 22 mJ/m$^2$. In the present alloy, the stacking fault energy is even higher, at 27 mJ/m$^2$ [56]. Yet, the stacking fault energy was reported to be lowered by approx. 40% when austenitic stainless steel contained approx. 270 ppm of hydrogen [57,58]. However, the stacking fault energy is not the only parameter influencing cell formation, a low shear modulus was also shown to play an important role [59]. Gavriljuk et al. [60] indicate that hydrogen reduces the shear modulus in austenitic steels, and they estimated that this can be up to over 30% [61]. In combination, these effects could reduce the activation energy for cross-slip and further promote cell formation.

However, we propose an alternative interpretation here. Hydrogen segregated at grain boundaries, which are typical sources of dislocations [45,46], can delay the emission of dislocations. Our suggestion is that partial locking by hydrogen of dislocations in or near the grain boundaries might decrease the segment length of dislocation sources in the boundary. Therefore, a higher activation stress is required. Without such locking effect (no hydrogen) the activation of dislocation sources with the highest Schmid-factor (i.e. single slip on the primary slip system or co-planar slip on the first two activated systems) is expected at the beginning of plastic deformation. Such co-planar slip is observed in the uncharged sample regions (Figures 5b, 5e, 5h). However, hydrogen locking, together with the higher stress level might favor the simultaneous activation of multiple slip systems already at the upper yield stress. The entanglement of these dislocations from multiple slip systems at an early stage of deformation, leads to the formation of cell-like structures (Figure 5a), which are, however, not related to the cell structures observed in materials with high stacking fault energy. This means that the early occurrence of cells might be not due to a faster dynamic recovery process because of an enhancement of the cross-slip frequency, but more a geometrical cell formation due to early multi-slip and entangling of dislocations of different slip systems. This interpretation does not require invoking an effect of hydrogen on the stacking fault energy or on the shear modulus, which it likely has but it is extremely challenging to quantify independently from other parameters. This interpretation is hence attractive considering the complexity of the problem at stake.

APT experiments confirm that there is no deuterium segregated to dislocations which is expected at the high tensile strain rate used in the current study. At the strain rate of $10^{-3}$ s$^{-1}$ and due to the low diffusivity of hydrogen in austenite, there is no interaction of hydrogen with dislocations because the velocity of dislocations is high which makes it difficult for hydrogen to follow the fast-moving dislocations. It hence enables us to study the dislocation structure evolved primarily due to the strengthening effect of hydrogen.

The dislocation structures in grains previously imaged by ECCI were examined again after approx. 2 months and none of the observed dislocations, stacking faults, dislocation partials and loop dipoles had changed (Figure S8). Previous reports of the evolution of dislocations moving during hydrogen desorption in Fe-Mn-alloys [18,42,62], could indicate that in our present observations, the hydrogen has not completely left the sample by the time the APT analysis was performed.

*4.3 5–7% deformation*
At 5% tensile strain, the continuous supply of more dislocations, along with dislocations extracted from the cell walls progressively have homogenized. The cell walls are not very dense and the force to extract dislocations from these cells or cell-like structures is likely low. The homogenization of the dislocation structure can lead to local variations in the hydrogen concentration via stress-assisted diffusion, as suggested by [63], causing a redistribution of the hydrogen. The local variations in hydrogen concentration can establish local variations in the stacking-fault energy, that can facilitate the nucleation of stacking faults [64,65] in the regions with relatively lower stacking fault energy, while dislocation loop dipoles are observed in other regions.

At higher tensile strain, i.e. 7%, relatively thin ε-martensite plates are observed. Keeping in mind that the stacking fault energy (γ) can be expressed as volume energy and surface energy contributions (assuming that the stacking fault is a two-atomic layer volume of hcp crystal structure with two surfaces on either side) [65]:

$$\gamma = \underbrace{n\rho_A(\Delta G^{chem} + E^{str})}_{\text{Volume energy term}} + \underbrace{2\sigma(n)}_{\text{Surface energy term}}$$

where $n$ is the number of fault planes, $\rho_A$ is the density of atoms in a close packed plane in moles per unit area, $\Delta G^{chem}$ and $E^{str}$ are respectively the chemical free energy difference between parent and product phases, and the strain energy, both defined as molar quantities, and $\sigma(n)$ is the surface energy per unit area of the particle/matrix interface.

Based on the suggestions by Koyama and coworkers [66], hydrogen stabilizes austenite and increases the volume energy contribution required to form the ε-martensite in austenitic steels. However, hydrogen promotes the formation of strain-induced ε-martensite [62,67], which suggests that the surface energy ($\sigma(n)$ in the equation above) is lowered by hydrogen, thereby facilitating the formation of ε-martensite. The ε-martensite formed in the studied TWIP steel is strain-induced [68], it is not observed in the uncharged region of the intermediate tensile strained (i.e., 3%, 5% and 7%) samples, however, it is observed in the uncharged sample deformed to tensile fracture, Figure S4. Hydrogen enables the formation of strain-induced ε-martensite at lower strain level, based on recent studies [67].

We, here, observe the formation of ε-martensite at 7% strain in the hydrogen-charged region which begins by the nucleation of stacking faults at 5% tensile strain. This nucleation of stacking faults leads to the formation of long, extended stacking faults which stack together to nucleate ε-martensite, which is in agreement with previous reports in austenitic steels [28,58]. The microstructure of the 5% tensile strained sample hence is a transient between the formation of entangled dislocations (forming cells because of the interaction of non-coplanar slip systems), and the formation of stacking faults.

Atomistic calculations suggest that hydrogen does not reside in the stacking fault in FCC iron containing carbon [69] because the presence of carbon atoms near the octahedral sites on the stacking fault plane makes the hydrogen segregation unfavorable. The tendency for hydrogen not to segregate to stacking faults could explain why we could not detect it by APT.

*4.4 10% deformation – fracture surface*
The intergranular fracture surface in the hydrogen-charged sample deformed to tensile fracture, Figure 4(b), could be explained by the thermodynamic theory of decohesion by Hirth and Rice [70]. It is beyond the scope of current study to quantify the hydrogen content leading to the decohesion of grain boundaries, yet we report hydrogen segregation by APT at a grain boundary from the sample deformed to 3%. Hydrogen segregated at grain boundaries of the sample deformed to fracture at 10% strain would reduce the critical fracture stress at the boundary which means that the strain on the sample at which grain boundary decohesion occurs is generally also reduced, following the proposed the HEDE mechanism [71], leading to the material's fracture [72].

## 5 Conclusion

We studied the dislocation structure resulting from the strengthening effect of hydrogen in a Fe-26.9Mn-0.28C (wt.%) TWIP steel. We observed a drop in ductility from 75% to 10% (as engineering strain to fracture) primarily due to the hardening of material by the presence of hydrogen, which is revealed by a 20% increase in the yield strength after hydrogen charging. The dislocation structures were examined at three different tensile strain levels enabling interpretation of the formation and dissolution of cell-like structures, followed by the formation of ε-martensite in the hydrogen-charged region. The cell-like structures are likely formed by the simultaneous interaction of several slip systems at early deformation stage, rather than by enhanced cross slip. The formation of strain-induced ε-martensite can be ascribed to a reduction in the martensite-austenite interface energy by hydrogen. This contrasts with the uncharged region where entangled dislocations are imaged at each tensile strain level. We conclude that the observed evolution of dislocation structures from the dislocation cell formation to the ε-martensite phase formation with an increase in the tensile strain is due to hydrogen. It was found that for the low strain observed here, the deformation structures were independent of Taylor factor and orientation of the principal strain axis.

The current study suggests that hydrogen segregated at grain boundaries, confirmed by deuterium-labelled APT analysis, delays the dislocation nucleation and hence hardens the material.

## Acknowledgements

The authors would like to thank Ms. Monika Nellessen, Ms. Katja Angenendt and Mr. Christian Bross for their support to the metallography sample preparation lab and SEM facilities at MPIE. We thank Uwe Tezins and Andreas Sturm for their technical support to the FIB and APT facilities at MPIE. Dr. Leigh T. Stephenson is gratefully acknowledged for helping with setting up the hydrogen charging facility. The authors would like to thank Prof. Dierk Raabe for helpful discussions. We also thank Mr. Michael Adamek for his help with the tensile tests, Mr. Frank Schlüter for his support at EDM facility and Mr. Eberhard Heinen for his help with the TDS experiments. H.K. and B.G. acknowledge the financial support from the ERC-CoG-SHINE-771602.

## References:

[1] J. Min, J. Lin, B. Sun, Effect of strain rate on spatio-temporal behavior of Portevin-Le Châtelier bands in a twinning induced plasticity steel, Mech. Mater. 68 (2014) 164–175. https://doi.org/10.1016/j.mechmat.2013.09.002.

[2] B.C. De Cooman, K. Chin, J. Kim, High Mn TWIP Steels for Automotive Applications, New Trends Dev. Automot. Syst. Eng. (2011). https://doi.org/10.5772/14086.

[3] B.C. De Cooman, O. Kwon, K.G. Chin, State-of-the-knowledge on TWIP steel, Mater. Sci. Technol. 28 (2012) 513–527. https://doi.org/10.1179/1743284711Y.0000000095.

[4] G. Dini, R. Ueji, A. Najafizadeh, S.M. Monir-Vaghefi, Flow stress analysis of TWIP steel via the XRD measurement of dislocation density, Mater. Sci. Eng. A. 527 (2010) 2759–2763.

[5] D.M. Bastidas, J. Ress, J. Bosch, U. Martin, Corrosion mechanisms of high-mn twinning-induced plasticity (Twip) steels: A critical review, Metals (Basel). 11 (2021) 1–45. https://doi.org/10.3390/met11020287.

[6] X. Guo, A. Schwedt, S. Richter, W. Bleck, Effects of Al on delayed fracture in TWIP steels – discussion from the aspects of structure homogeneity, hydrogen traps and corrosion resistance, SteelyHydrogen2014 Conf. Proc. (2014) 5–7.

[7] H.K.D.H. Bhadeshia, Prevention of Hydrogen Embrittlement in Steels, ISIJ Int. 56 (2016) 24–36. https://doi.org/10.2355/isijinternational.ISIJINT-2015-430.

[8] M. Koyama, E. Akiyama, K. Tsuzaki, Hydrogen embrittlement in a Fe-Mn-C ternary twinning-induced plasticity steel, Corros. Sci. 54 (2012) 1–4. https://doi.org/10.1016/j.corsci.2011.09.022.

[9] M. Koyama, E. Akiyama, K. Tsuzaki, D. Raabe, Hydrogen-assisted failure in a twinning-induced plasticity steel studied under in situ hydrogen charging by electron channeling contrast imaging, Acta Mater. 61 (2013) 4607–4618. https://doi.org/10.1016/j.actamat.2013.04.030.

[10] B. Bal, M. Koyama, G. Gerstein, H.J. Maier, K. Tsuzaki, Effect of strain rate on hydrogen embrittlement susceptibility of twinning-induced plasticity steel pre-charged with high-pressure hydrogen gas, Int. J. Hydrogen Energy. 41 (2016) 15362–15372. https://doi.org/10.1016/j.ijhydene.2016.06.259.

[11] M. Koyama, E. Akiyama, K. Tsuzaki, Hydrogen-induced delayed fracture of a Fe-22Mn-0.6C steel pre-strained at different strain rates, Scr. Mater. 66 (2012) 947–950. https://doi.org/10.1016/j.scriptamat.2012.02.040.

[12] Q. Liu, A. Atrens, A critical review of the influence of hydrogen on the mechanical properties of medium-strength steels, Corros. Rev. 31 (2013) 85–103. https://doi.org/10.1515/corrrev-2013-0023.

[13] Y. Ogawa, H. Hosoi, K. Tsuzaki, T. Redarce, O. Takakuwa, H. Matsunaga, Hydrogen, as an alloying element, enables a greater strength-ductility balance in an Fe-Cr-Ni-based, stable austenitic stainless steel, Acta Mater. 199 (2020) 181–192. https://doi.org/10.1016/j.actamat.2020.08.024.

[14] T. Hickel, R. Nazarov, E.J. McEniry, G. Leyson, B. Grabowski, J. Neugebauer, Ab initio based understanding of the segregation and diffusion mechanisms of hydrogen in steels, Jom. 66 (2014) 1399–1405. https://doi.org/10.1007/s11837-014-1055-3.

[15] B. Sun, D. Wang, X. Lu, D. Wan, D. Ponge, X. Zhang, Current Challenges and Opportunities Toward Understanding Hydrogen Embrittlement Mechanisms in Advanced High-Strength Steels: A Review, Acta Metall. Sin. (English Lett. 34 (2021) 741–754. https://doi.org/10.1007/s40195-021-01233-1.

[16] T. Neeraj, R. Srinivasan, J. Li, Hydrogen embrittlement of ferritic steels: Observations on deformation microstructure, nanoscale dimples and failure by nanovoiding, Acta Mater. 60 (2012) 5160–5171. https://doi.org/10.1016/j.actamat.2012.06.014.

[17] X. Lu, D. Wang, D. Wan, Z.B. Zhang, N. Kheradmand, A. Barnoush, Effect of electrochemical charging on the hydrogen embrittlement susceptibility of alloy 718, Acta Mater. 179 (2019) 36–48. https://doi.org/10.1016/j.actamat.2019.08.020.

[18] M. Koyama, S.M. Taheri-Mousavi, S.M. Taheri-Mousavi, H. Yan, J. Kim, B.C. Cameron, S.S. Moeini-Ardakani, J. Li, J. Li, C.C. Tasan, Origin of micrometer-scale dislocation motion during hydrogen desorption, Sci. Adv. 6 (2020). https://doi.org/10.1126/sciadv.aaz1187.

[19]   S. Zaefferer, N.-N. Elhami, Theory and application of electron channelling contrast imaging under controlled diffraction conditions, Acta Mater. 75 (2014) 20–50. https://doi.org/10.1016/j.actamat.2014.04.018.

[20]   B. et al Gault, A. Chiaramonti, O. Cojocaru-Mirédin, P. Stender, R. Dubosq, C. Freysoldt, S.K. Makineni, T. Li, M. Moody, J.M. Cairney, Atom Probe Tomography, Nat. Rev. Methods Prim. (2021) 1–51.

[21]   Y.S. Chen, H. Lu, J. Liang, A. Rosenthal, H. Liu, G. Sneddon, I. McCarroll, Z. Zhao, W. Li, A. Guo, J.M. Cairney, Observation of hydrogen trapping at dislocations, grain boundaries, and precipitates, Science (80-. ). 367 (2020) 171–175. https://doi.org/10.1126/science.aaz0122.

[22]   J. Takahashi, K. Kawakami, T. Tarui, Direct observation of hydrogen-trapping sites in vanadium carbide precipitation steel by atom probe tomography, Scr. Mater. 67 (2012) 213–216. https://doi.org/10.1016/j.scriptamat.2012.04.022.

[23]   J. Takahashi, K. Kawakami, Y. Kobayashi, Origin of hydrogen trapping site in vanadium carbide precipitation strengthening steel, Acta Mater. 153 (2018) 193–204. https://doi.org/10.1016/j.actamat.2018.05.003.

[24]   I.E. McCarroll, Y.-C. Lin, A. Rosenthal, H.-W. Yen, J.M. Cairney, Hydrogen trapping at dislocations, carbides, copper precipitates and grain boundaries in a dual precipitating low-carbon martensitic steel, Scr. Mater. 221 (2022) 114934. https://doi.org/10.1016/J.SCRIPTAMAT.2022.114934.

[25]   M. Daamen, S. Richter, G. Hirt, Microstructure analysis of high-manganese TWIP steels produced via strip casting, Key Eng. Mater. 554–557 (2013) 553–561. https://doi.org/10.4028/www.scientific.net/KEM.554-557.553.

[26]   D. Pérez Escobar, L. Duprez, A. Atrens, K. Verbeken, Influence of experimental parameters on thermal desorption spectroscopy measurements during evaluation of hydrogen trapping, J. Nucl. Mater. 450 (2014) 32–41. https://doi.org/10.1016/j.jnucmat.2013.07.006.

[27]   X. Zhu, W. Li, T.Y. Hsu, S. Zhou, L. Wang, X. Jin, Improved resistance to hydrogen embrittlement in a high-strength steel by quenching-partitioning-tempering treatment, Scr. Mater. 97 (2015) 21–24. https://doi.org/10.1016/j.scriptamat.2014.10.030.

[28]   N. Narita, C.J. Altstetter, H.K. Birnbaum, Hydrogen-Related Phase Transformations in Austenitic Stainless Steels., Metall. Trans. A, Phys. Metall. Mater. Sci. 13 A (1982) 1355–1365. https://doi.org/10.1007/BF02642872.

[29]   X. Guo, S. Zaefferer, F. Archie, W. Bleck, Hydrogen effect on the mechanical behaviour and microstructural features of a Fe-Mn-C twinning induced plasticity steel, Int. J. Miner. Metall. Mater. 28 (2021) 835–846. https://doi.org/10.1007/s12613-021-2284-4.

[30]   N. Zan, H. Ding, X. Guo, Z. Tang, W. Bleck, Effects of grain size on hydrogen embrittlement in a Fe-22Mn-0.6C TWIP steel, Int. J. Hydrogen Energy. 40 (2015) 10687–10696. https://doi.org/10.1016/j.ijhydene.2015.06.112.

[31]   H. Khanchandani, S.-H. Kim, R.S. Varanasi, T. Prithiv, L.T. Stephenson, B. Gault, Hydrogen and deuterium charging of lifted-out specimens for atom probe tomography, Open Res. Eur. 1 (2022) 122. https://doi.org/10.12688/openreseurope.14176.2.

[32]   S. Zaefferer, G. Habler, Scanning electron microscopy and electron backscatter diffraction, Eur. Mineral. Union Notes Mineral. 16 (2017) 37–95. https://doi.org/10.1180/EMU-notes.16.3.

[33]   M.K. Miller, K.F. Russell, K. Thompson, R. Alvis, D.J. Larson, Review of atom probe FIB-based specimen preparation methods, Microsc. Microanal. 13 (2007) 428–436. https://doi.org/10.1017/S1431927607070845.

[34] B. Sun, W. Lu, B. Gault, R. Ding, S.K. Makineni, D. Wan, C.-H. Wu, H. Chen, D. Ponge, D. Raabe, Chemical heterogeneity enhances hydrogen resistance in high-strength steels, Nat. Mater. 2021. (2021) 1–6. https://doi.org/10.1038/s41563-021-01050-y.

[35] A.J. Breen, L.T. Stephenson, B. Sun, Y. Li, O. Kasian, D. Raabe, M. Herbig, B. Gault, Solute hydrogen and deuterium observed at the near atomic scale in high-strength steel, Acta Mater. 188 (2020) 108–120. https://doi.org/10.1016/j.actamat.2020.02.004.

[36] Y.H. Chang, I. Mouton, L. Stephenson, M. Ashton, G.K. Zhang, A. Szczpaniak, W.J. Lu, D. Ponge, D. Raabe, B. Gault, Quantification of solute deuterium in titanium deuteride by atom probe tomography with both laser pulsing and high-voltage pulsing: Influence of the surface electric field, New J. Phys. 21 (2019) 053025. https://doi.org/10.1088/1367-2630/ab1c3b.

[37] H. Khanchandani, A.A. El-Zoka, S.-H. Kim, U. Tezins, D. Vogel, A. Sturm, D. Raabe, B. Gault, L. Stephenson, Laser-equipped gas reaction chamber for probing environmentally sensitive materials at near atomic scale, PLoS One. (2021) 1–19. https://doi.org/10.1371/journal.pone.0262543.

[38] L.T. Stephenson, A. Szczepaniak, I. Mouton, K.A.K. Rusitzka, A.J. Breen, U. Tezins, A. Sturm, D. Vogel, Y. Chang, P. Kontis, A. Rosenthal, J.D. Shepard, U. Maier, T.F. Kelly, D. Raabe, B. Gault, The Laplace project: An integrated suite for preparing and transferring atom probe samples under cryogenic and UHV conditions, PLoS One. 13 (2018) 1–13. https://doi.org/10.1371/journal.pone.0209211.

[39] T. Kanezaki, C. Narazaki, Y. Mine, S. Matsuoka, Y. Murakami, Effects of hydrogen on fatigue crack growth behavior of austenitic stainless steels, Int. J. Hydrogen Energy. 33 (2008) 2604–2619. https://doi.org/10.1016/j.ijhydene.2008.02.067.

[40] M. Koyama, S. Okazaki, T. Sawaguchi, K. Tsuzaki, Hydrogen Embrittlement Susceptibility of Fe-Mn Binary Alloys with High Mn Content: Effects of Stable and Metastable ε-Martensite, and Mn Concentration, Metall. Mater. Trans. A Phys. Metall. Mater. Sci. 47 (2016) 2656–2673. https://doi.org/10.1007/s11661-016-3431-9.

[41] C. Zhang, H. Zhi, S. Antonov, L. Chen, Y. Su, Hydrogen-enhanced densified twinning (HEDT) in a twinning-induced plasticity (TWIP) steel, Scr. Mater. 190 (2021) 108–112. https://doi.org/10.1016/j.scriptamat.2020.08.047.

[42] D. An, X. Zhang, S. Zaefferer, The combined and interactive effects of orientation, strain amplitude, cycle number, stacking fault energy and hydrogen doping on microstructure evolution of polycrystalline high-manganese steels under low-cycle fatigue, Int. J. Plast. 134 (2020) 102803. https://doi.org/10.1016/j.ijplas.2020.102803.

[43] H. Idrissi, K. Renard, L. Ryelandt, D. Schryvers, P.J. Jacques, On the mechanism of twin formation in Fe–Mn–C TWIP steels, Acta Mater. 58 (2010) 2464–2476. https://doi.org/10.1016/J.ACTAMAT.2009.12.032.

[44] E.I. Galindo-Nava, P.E.J. Rivera-Díaz-del-Castillo, Understanding martensite and twin formation in austenitic steels: A model describing TRIP and TWIP effects, Acta Mater. 128 (2017) 120–134. https://doi.org/10.1016/J.ACTAMAT.2017.02.004.

[45] T. Shimokawa, Asymmetric ability of grain boundaries to generate dislocations under tensile or compressive loadings, Phys. Rev. B - Condens. Matter Mater. Phys. 82 (2010) 1–13. https://doi.org/10.1103/PhysRevB.82.174122.

[46] C.Y. Hung, Y. Bai, T. Shimokawa, N. Tsuji, M. Murayama, A correlation between grain boundary character and deformation twin nucleation mechanism in coarse-grained high-Mn austenitic steel, Sci. Rep. 11 (2021) 1–13. https://doi.org/10.1038/s41598-021-87811-w.

[47] D.R. Kingham, The post-ionization of field evaporated ions: A theoretical explanation of

multiple charge states, Surf. Sci. 116 (1982) 273–301. https://doi.org/10.1016/0039-6028(82)90434-4.

[48] B. Gault, M.P. Moody, J.M. Cairney, P.R. Simon, Atom probe microscopy, 2012.

[49] M.P. Moody, L.T. Stephenson, A. V. Ceguerra, S.P. Ringer, Quantitative binomial distribution analyses of nanoscale like-solute atom clustering and segregation in atom probe tomography data, Microsc. Res. Tech. 71 (2008) 542–550. https://doi.org/10.1002/jemt.20582.

[50] L.T. Stephenson, A. V. Ceguerra, T. Li, T. Rojhirunsakool, S. Nag, R. Banerjee, J.M. Cairney, S.P. Ringer, Point-by-point compositional analysis for atom probe tomography, MethodsX. 1 (2014) 12–18. https://doi.org/10.1016/j.mex.2014.02.001.

[51] M. Müller, G.D.W.D.W.D.W. Smith, B. Gault, C.R.M.R.M.R.M. Grovenor, Phase separation in thick InGaN layers – A quantitative, nanoscale study by pulsed laser atom probe tomography, Acta Mater. 60 (2012) 4277–4285. https://doi.org/10.1016/j.actamat.2012.04.030.

[52] D.K.Y. Tam, S. Ruan, P. Gao, T. Yu, High-performance ballistic protection using polymer nanocomposites, Adv. Mil. Text. Pers. Equip. (2012) 213–237. https://doi.org/10.1533/9780857095572.2.213.

[53] R. Silverstein, D. Eliezer, B. Glam, Hydrogen Effect on Duplex Stainless Steels at Very High Strain Rates, Energy Procedia. 107 (2017) 199–204. https://doi.org/10.1016/j.egypro.2016.12.172.

[54] E.G. Astafurova, G.G. Zakharova, H.J. Maier, Hydrogen-induced twinning in ⟨0 0 1⟩ Hadfield steel single crystals, Scr. Mater. 63 (2010) 1189–1192. https://doi.org/10.1016/j.scriptamat.2010.08.029.

[55] I. Gutierrez-Urrutia, D. Raabe, Dislocation and twin substructure evolution during strain hardening of an Fe-22 wt.% Mn-0.6 wt.% C TWIP steel observed by electron channeling contrast imaging, Acta Mater. 59 (2011) 6449–6462. https://doi.org/10.1016/j.actamat.2011.07.009.

[56] A. Saeed-Akbari, J. Imlau, U. Prahl, W. Bleck, Derivation and variation in composition-dependent stacking fault energy maps based on subregular solution model in high-manganese steels, Metall. Mater. Trans. A Phys. Metall. Mater. Sci. 40 (2009) 3076–3090. https://doi.org/10.1007/s11661-009-0050-8.

[57] A.E. Pontini, J.D. Hermida, X-ray diffraction measurement of the stacking fault energy reduction induced by hydrogen in an AISI 304 steel, Scr. Mater. 37 (1997) 1831–1837. https://doi.org/10.1016/S1359-6462(97)00332-1.

[58] J.D. Hermida, A. Roviglione, Stacking fault energy decrease in austenitic stainless steels induced by hydrogen pairs formation, Scr. Mater. 39 (1998) 1145–1149. https://doi.org/10.1016/S1359-6462(98)00285-1.

[59] P. Landau, R.Z. Shneck, G. Makov, A. Venkert, Evolution of dislocation patterns in fcc metals, IOP Conf. Ser. Mater. Sci. Eng. 3 (2009). https://doi.org/10.1088/1757-899X/3/1/012004.

[60] V.G. Gavriljuk, V.N. Shivanyuk, B.D. Shanina, Change in the electron structure caused by C, N and H atoms in iron and its effect on their interaction with dislocations, Acta Mater. 53 (2005) 5017–5024. https://doi.org/10.1016/j.actamat.2005.07.028.

[61] V.G. Gavriljuk, B.D. Shanina, V.N. Shyvanyuk, S.M. Teus, Electronic effect on hydrogen brittleness of austenitic steels, J. Appl. Phys. 108 (2010) 083723. https://doi.org/10.1063/1.3499610.

[62] D. An, W. Krieger, S. Zaefferer, Unravelling the effect of hydrogen on microstructure

evolution under low-cycle fatigue in a high-manganese austenitic TWIP steel, Int. J. Plast. (2019). https://doi.org/10.1016/j.ijplas.2019.11.004.

[63] J.P. Chateau, D. Delafosse, T. Magnin, Numerical simulations of hydrogen–dislocation interactions in fcc stainless steels.: part I: hydrogen–dislocation interactions in bulk crystals, Acta Mater. 50 (2002) 1507–1522. https://doi.org/10.1016/S1359-6454(02)00008-3.

[64] L. Pizzagalli, J. Godet, S. Brochard, H.J. Gotsis, T. Albaret, Stacking fault formation created by plastic deformation at low temperature and small scales in silicon, Phys. Rev. Mater. 4 (2020). https://doi.org/10.1103/PhysRevMaterials.4.093603.

[65] G.B. Olson, M. Cohen, A general mechanism of martensitic nucleation: Part I. General Concepts and FCC -> HCP transformation, Metall. Trans. A. 7 (1976) 1897. https://doi.org/10.1007/BF02654989.

[66] M. Koyama, K. Hirata, Y. Abe, A. Mitsuda, S. Iikubo, K. Tsuzaki, An unconventional hydrogen effect that suppresses thermal formation of the hcp phase in fcc steels, Sci. Rep. 8 (2018) 1–7. https://doi.org/10.1038/s41598-018-34542-0.

[67] M. Koyama, N. Terao, K. Tsuzaki, Revisiting the effects of hydrogen on deformation-induced γ-ε martensitic transformation, Mater. Lett. 249 (2019) 197–200. https://doi.org/10.1016/j.matlet.2019.04.093.

[68] A. Talgotra, E. Nagy, M. Sepsi, M. Benke, V. Mertinger, Strain induced martensitic transformations in FeMnCr TWIP steel, IOP Conf. Ser. Mater. Sci. Eng. 426 (2018). https://doi.org/10.1088/1757-899X/426/1/012046.

[69] S. Simonetti, L. Moro, N.E. Gonzalez, G. Brizuela, A. Juan, Quantum chemical study of C and H location in an fcc stacking fault, Int. J. Hydrogen Energy. 29 (2004) 649–658. https://doi.org/10.1016/S0360-3199(03)00218-0.

[70] J.P. Hirth, J.R. Rice, On the thermodynamics of adsorption at interfaces as it influences decohesion, Metall. Trans. A. 11 (1980) 1501–1511. https://doi.org/10.1007/BF02654514.

[71] Y. Liang, P. Sofronis, Toward a phenomenological description of hydrogen-induced decohesion at particle/matrix interfaces, J. Mech. Phys. Solids. 51 (2003) 1509–1531. https://doi.org/10.1016/S0022-5096(03)00052-8.

[72] Y.A. Du, L. Ismer, J. Rogal, T. Hickel, J. Neugebauer, R. Drautz, First-principles study on the interaction of H interstitials with grain boundaries in α- and γ-Fe, Phys. Rev. B - Condens. Matter Mater. Phys. 84 (2011). https://doi.org/10.1103/PhysRevB.84.144121.

**SUPPLEMENTARY INFORMATION**

**1 Thermal desorption spectroscopy (TDS)**

In order to measure the hydrogen content of the samples before and after charging, TDS was carried out using a Hiden TDS workstation equipped with an HAL 3F 510 PIC quadrupole mass spectrometer. The TDS sample of dimensions $10 \times 15 \times 1$ mm$^3$ was first polished until the final mechano-chemical polishing step and then charged with hydrogen similarly by cathodic hydrogen charging for 5 days at ambient temperature conditions. The hydrogen-charged TDS sample was positioned in an ultra-high vacuum chamber and heated at a constant rate of 16°C/min from room temperature to 800°C. The heating leads to the desorption of hydrogen from the sample which was detected by a quadrupole mass spectrometer. A curve of hydrogen desorption rate (wt. ppm/s) vs. temperature (°C) was hence plotted. The area underneath the curve was determined to provide the total hydrogen content of the sample.

The hydrogen desorption rate (wt. ppm/s) vs. temperature (°C) curve obtained from TDS analysis of the hydrogen-charged and the uncharged samples is shown in Figure S1. The total hydrogen content in the sample was determined from the area underneath the curve as 13.2 wt. ppm in the hydrogen-charged sample (pink curve) against 1.3 wt. ppm in the uncharged sample (black curve). It must be noted that TDS experiments were performed on the recrystallized samples. The peak at ~120°C in the hydrogen-charged sample shows that diffusible hydrogen was introduced into the sample by cathodic charging, thereby confirming the charging method.

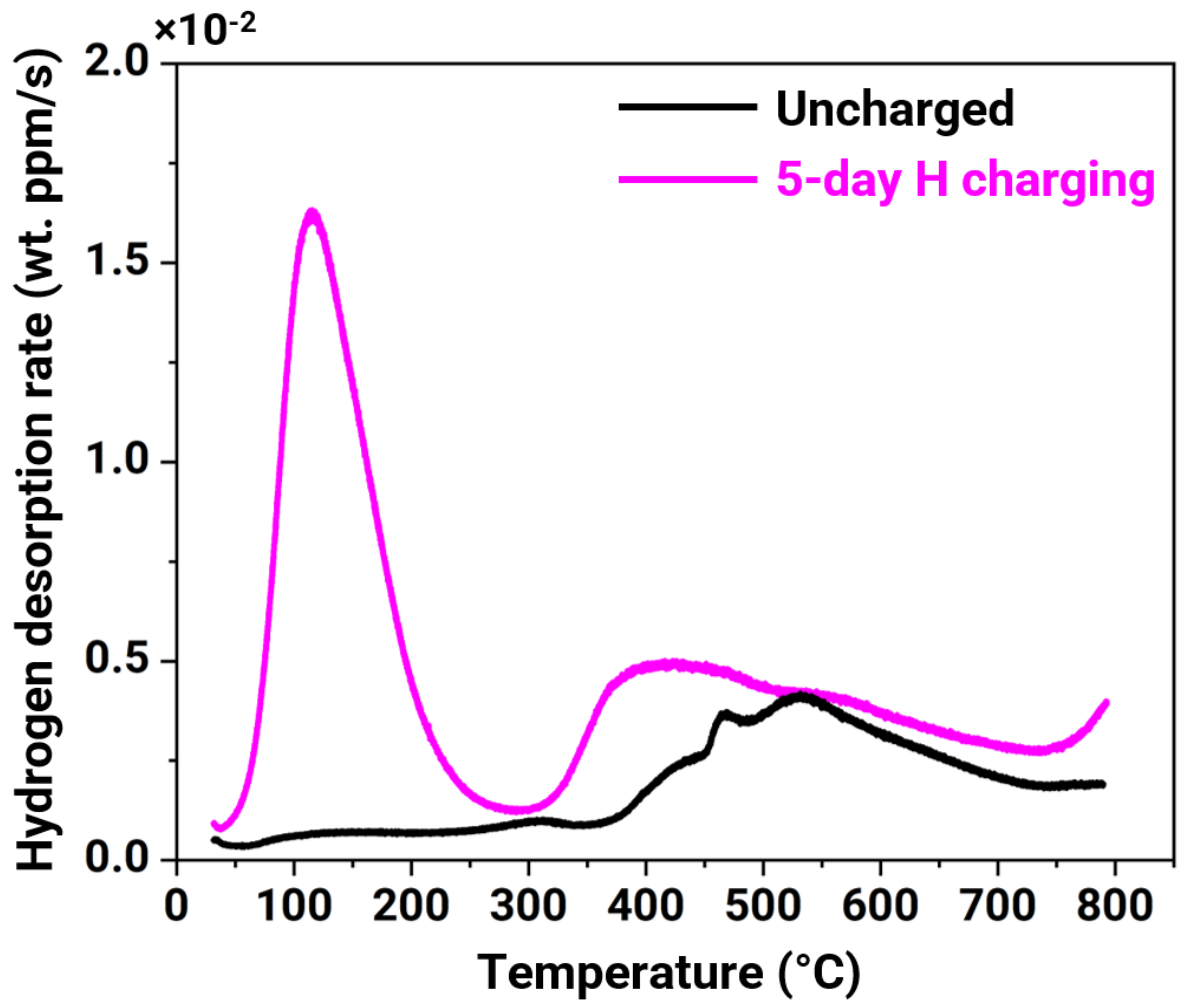

*Figure S1. The thermal desorption spectroscopy curve of the uncharged and the hydrogen-charged recrystallized samples.*

## 2 Energy dispersive X-ray spectroscopy (EDX) mapping

Figure S2(a) shows the SEM image of the region close to the side edges of the sample surface that was chosen for the EDX mapping, while Figures S2(b-c) show the iron and manganese distribution in that region respectively. Similarly, Figure S2(d) shows the SEM image of the center region from the same sample that was chosen for the EDX mapping, while Figures S2(e-f) show the iron and manganese distribution in that region respectively. Figure S2(g) displays the SEM image of the region that was chosen for a composition line profile from the surface to the center of the sample. The corresponding composition profile of iron and manganese is drawn in Figure S2(h), showing homogeneous composition from the surface to the center of the sample.

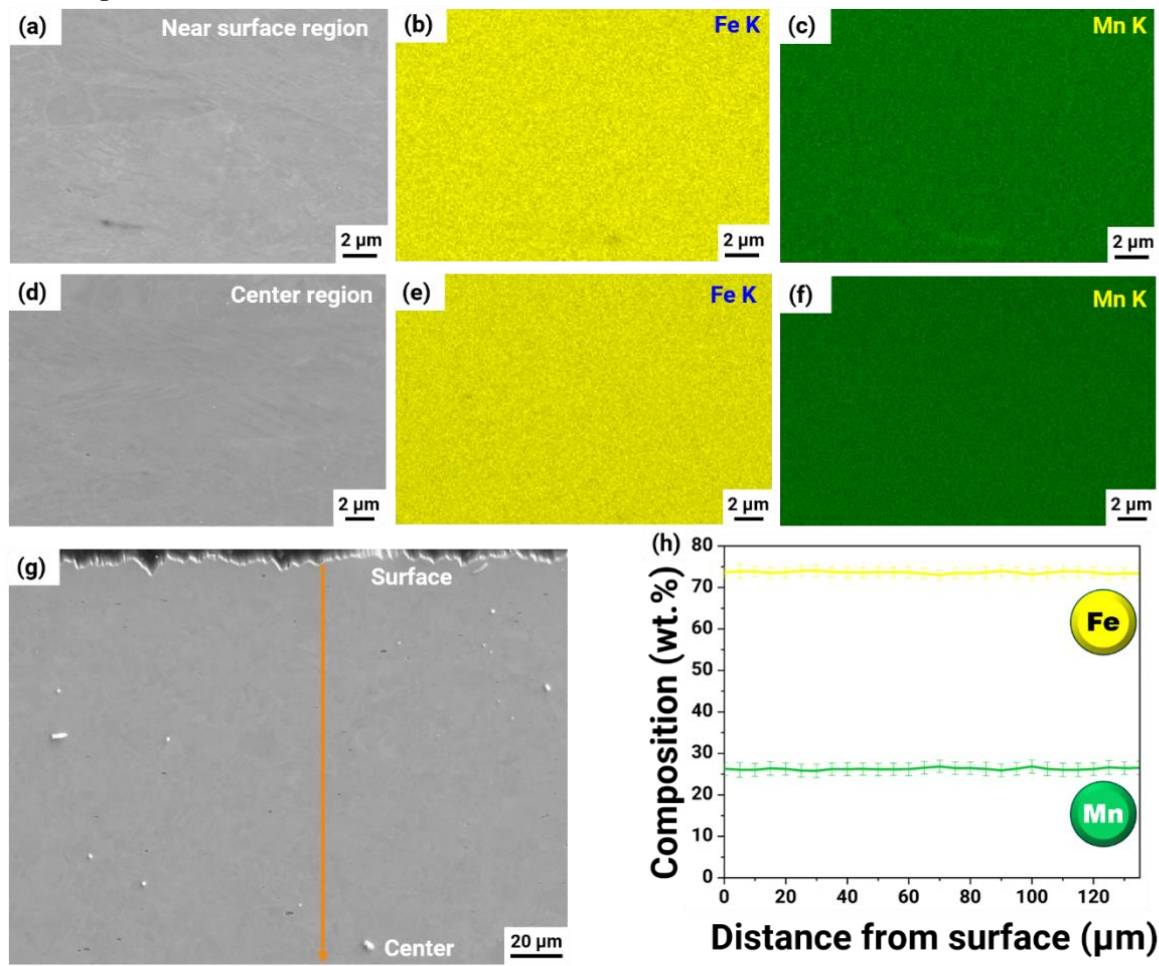

*Figure S2. (a) SEM image of the region chosen for the EDX mapping of (b) iron and (c) manganese near the surface of the sample; (d) SEM image of the region chosen for the EDX mapping of (e) iron and (f) manganese at the center of the sample; (g) SEM image of the region selected for the composition line profile shown in (h) from the surface to the center of the sample.*

## 3 Tensile test curves

The engineering stress-strain curves of two hydrogen-charged and two uncharged samples are shown in Figure S3.

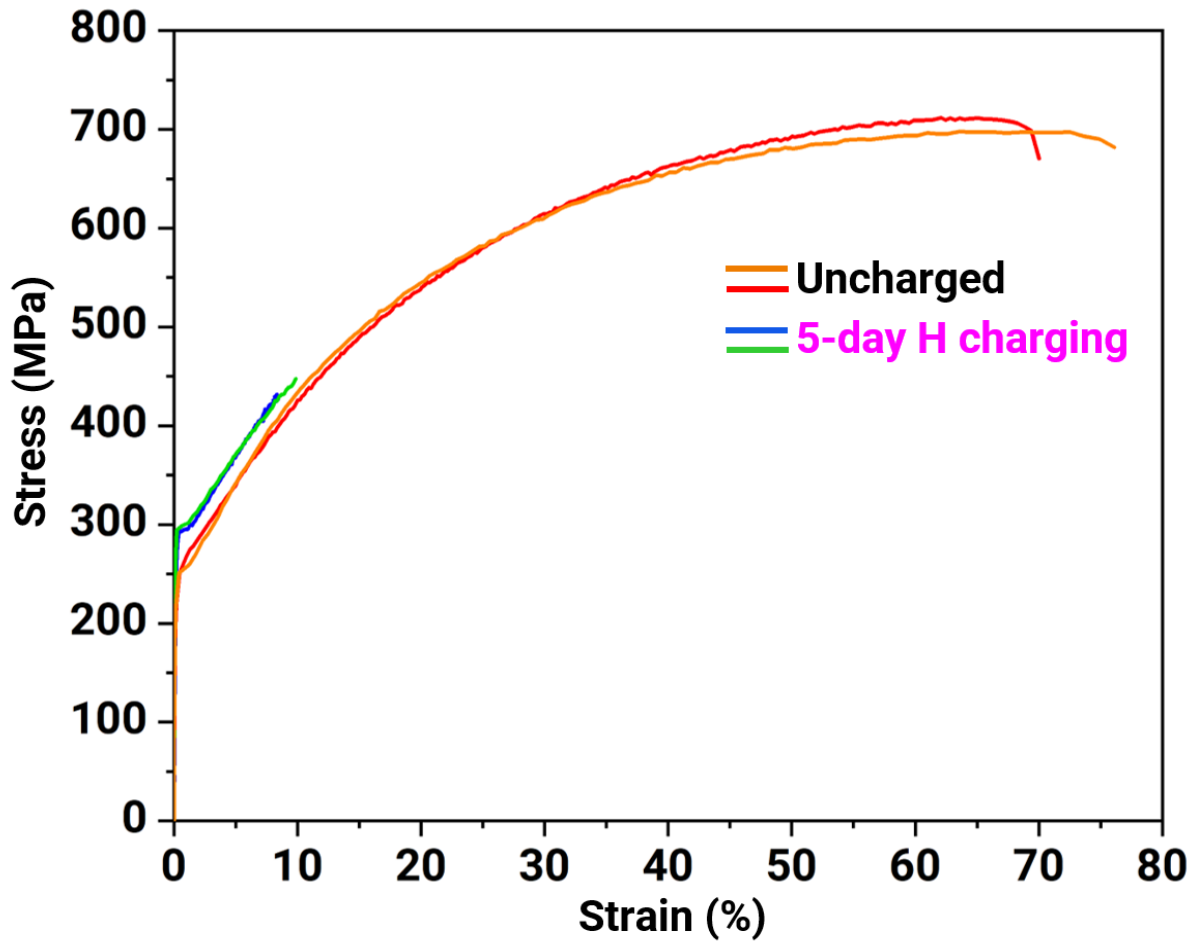

*Figure S3. The engineering stress-strain curves of the hydrogen-charged and the uncharged samples.*

**4 Electron-channeling contrast imaging of the samples deformed to tensile fracture**
ECC micrographs of the samples deformed to tensile fracture were observed from the region delineated by a blue box in the schematic shown in Figure 1(b), approx. 500 µm away from the fracture surface. Figure S4(a) shows an ECC micrograph of the hydrogen-charged sample deformed to tensile fracture, from a grain located in the hydrogen-charged region, i.e., within the ~100 µm distance from the sample surface. The plates of ε-martensite were observed along with dislocation tangles. Similar microstructural features were observed in the uncharged sample deformed to tensile fracture, as shown by an ECC micrograph in Figure S4(b). Note that the figures were taken from significantly differently strained samples. From these figures, it was hence very difficult to draw any conclusion about the deformation mechanism due to the presence of an extremely high and heterogeneous tensile strain in both samples deformed to tensile fracture. The dislocation structures were hence examined at intermediate levels of tensile strain, i.e., 3%, 5% and 7% to better capture the formation mechanism of the deformed microstructure.

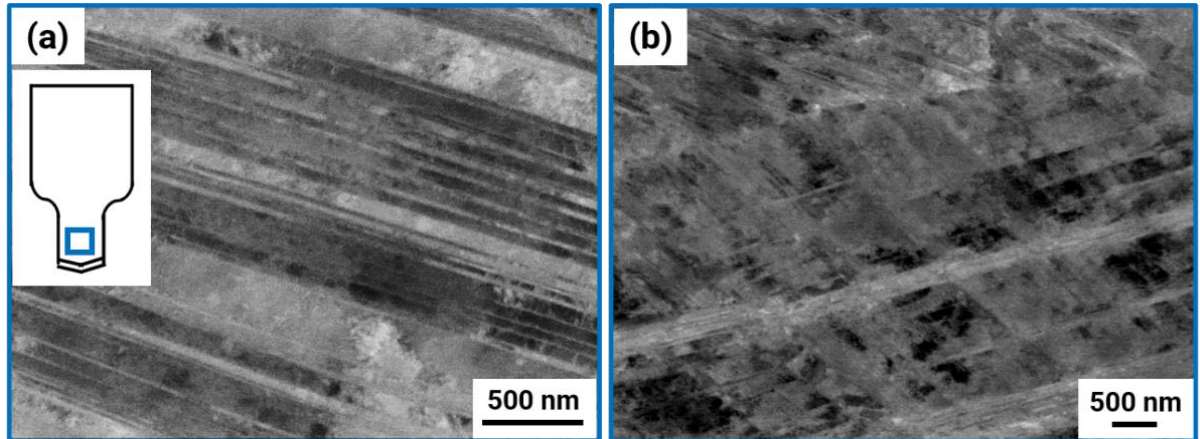

*Figure S4. (a) ECC micrograph of the hydrogen-charged sample deformed to tensile fracture (ε=10 %); and (b) ECC micrograph of the uncharged sample deformed to tensile fracture (ε=75 %); both images were examined from the region delineated by blue box in the schematic shown in Figure 1(b).*

## 5 Statistical analyses of the dislocation structure

The dislocation structure of approx. 20 randomly-selected grains was examined by ECCI from both the hydrogen-charged and the uncharged regions of each of the three samples, i.e., 3%, 5% and 7% tensile strained. Their EBSD mapping was performed and the corresponding inverse pole figure maps and Taylor factor maps of the investigated grains are shown in Figure S5. The apparent difference in grain size is in part related to the scale and inhomogeneous grain size distribution across the material – see Figure 2(b).

For a given amount of external strain, the total amount of shear accumulated in the active individual slip systems of a given grain depends on the crystal orientation and is quantified by the Taylor factor. A high Taylor factor indicates that a large amount of shear (i.e. a high density of dislocations) is required to accommodate the external strain. For the FCC crystal structure under uniaxial loading, the Taylor factor takes values between 2.2 and 3.6. The Taylor factors for tensile loading and activity of {111}<0 1 -1> slip systems were evaluated from the orientation data. The Taylor factor values of all the investigated grains lie in between 2.3 and 3.6, Figure S5, which indicate that the microstructural evolution observations were not biased by the local Taylor factor.

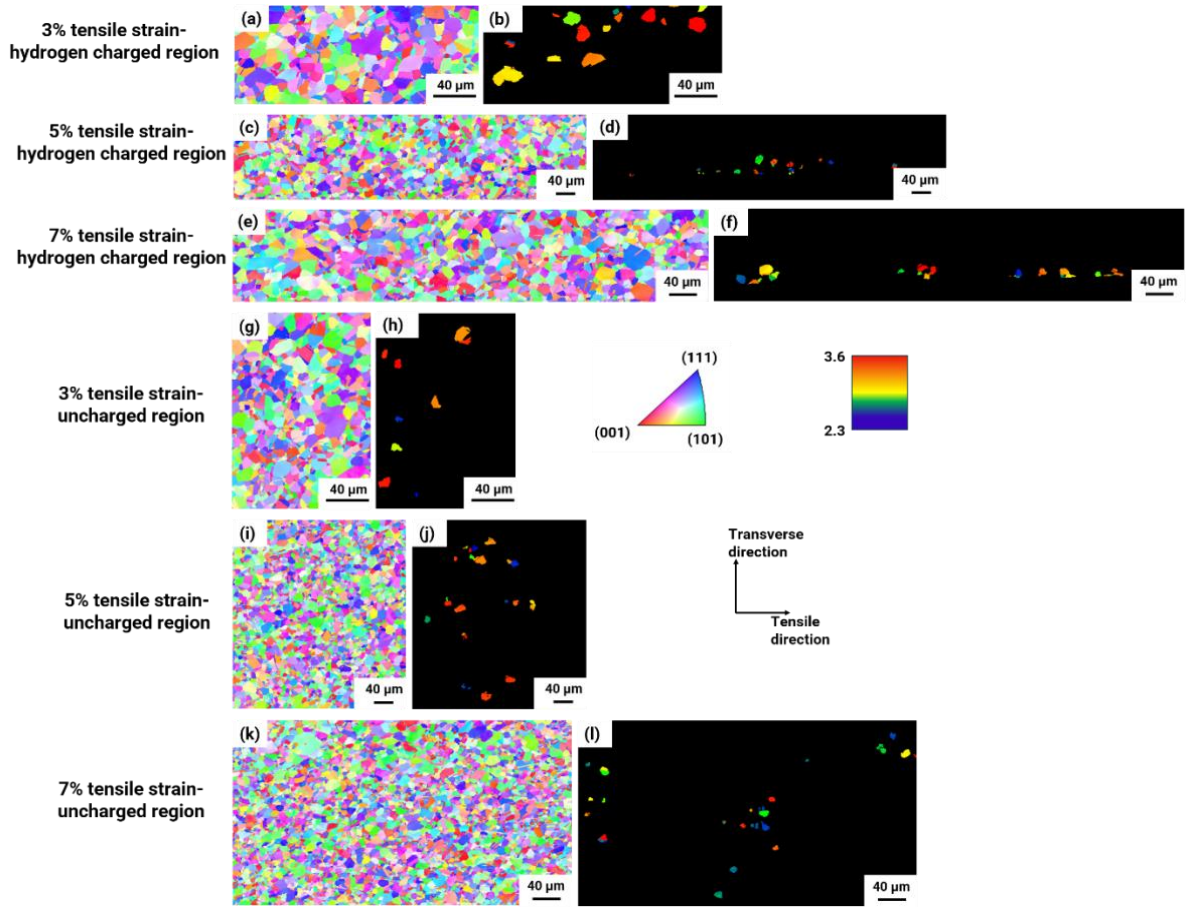

*Figure S5. EBSD-IPF mapping and the Taylor factor maps of the grains examined by correlative ECCI from the hydrogen-charged and the uncharged regions of each of the three samples, i.e., 3% tensile strained-hydrogen charged (a-b), 5% tensile strained-hydrogen charged (c-d), 7% tensile strained-hydrogen charged (e-f), 3% tensile strained-uncharged (g-h), 5% tensile strained-uncharged (i-j), 7% tensile strained-uncharged (k-l), respectively.*

The corresponding [010] inverse pole figures in the tensile direction are displayed in Figure S6.

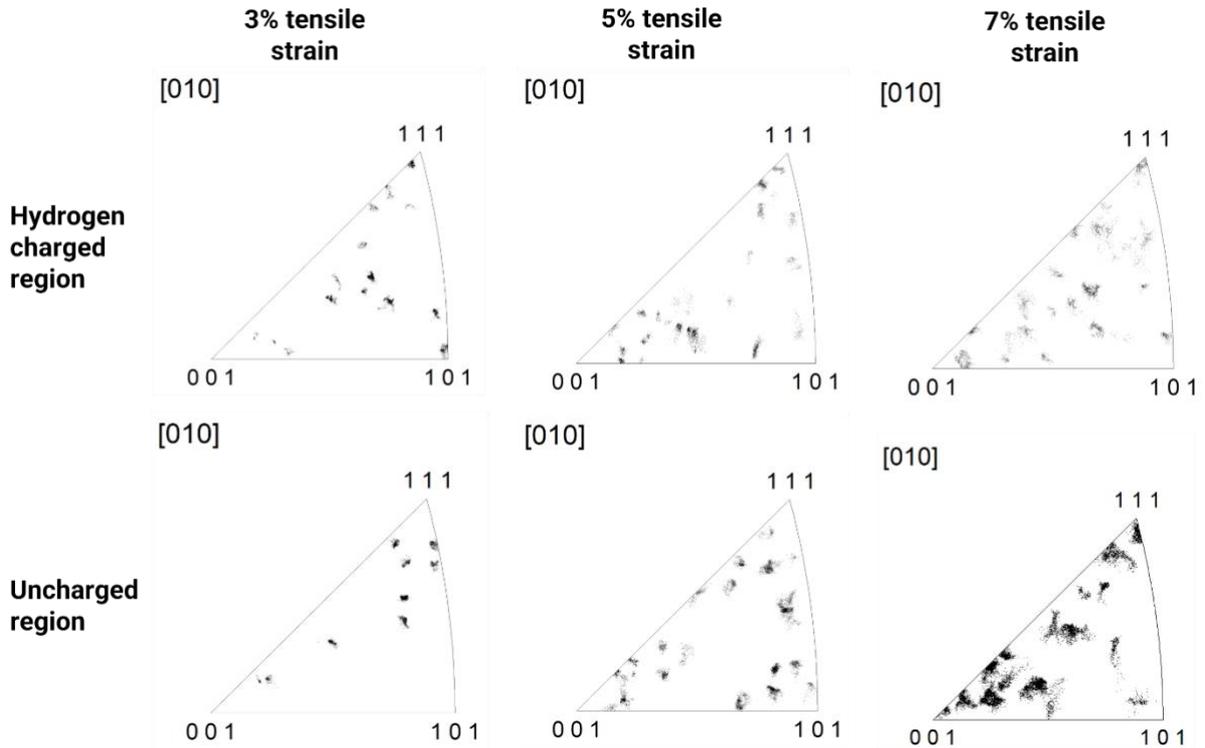

*Figure S6. The [010] inverse pole figures for each of the examined grains from the hydrogen-charged and the uncharged regions of each of the three samples, i.e., 3%, 5% and 7%.*

## 6 APT dataset of deuterium charged stacking faults

Figure S7(a) depicts the hydrogen-charged region from the 7% tensile strained sample containing the stacking faults. An APT specimen with stacking faults was charged with deuterium, whose 3-D elemental map is shown in Figure S7(b). No elemental segregation was observed at these stacking faults. The binomial frequency distribution analysis is shown in Figure S7(c), indicating the homogeneous distribution of deuterium across the APT dataset.

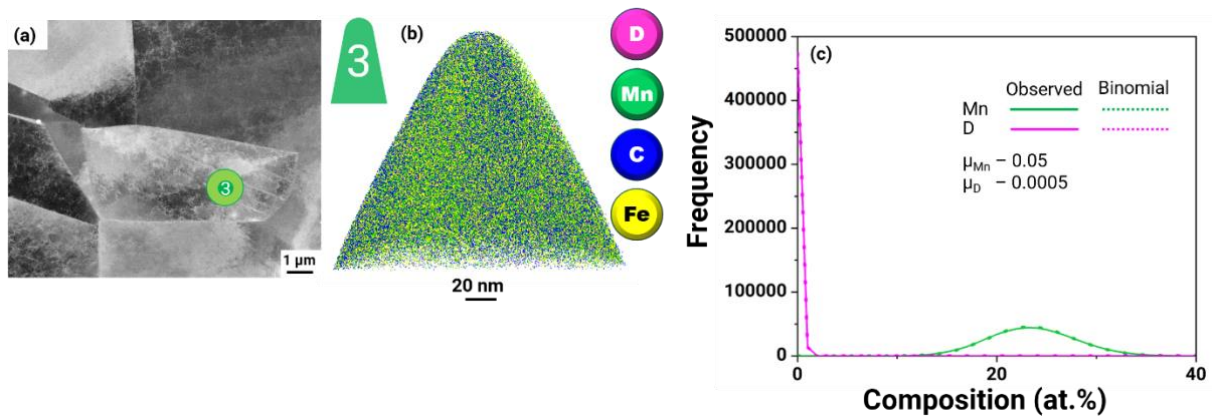

*Figure S7. (a) ECC image indicating the region in the hydrogen-charged region of the 7% tensile strained sample from where the APT specimen was prepared; (b) 3-D elemental map showing deuterium (D), manganese (Mn), carbon (C) and iron (Fe) atoms in the APT specimen containing stacking faults, whose binomial frequency distribution analysis is shown in (c).*

# 7 Stability of the dislocation structures

The dislocation structures were examined again after approx. 2 months in order to confirm their stability. Here, the dislocation cells formed in the hydrogen-charged region of the 3% tensile strained sample are shown as an example to prove that the microstructural changes associated to the presence of hydrogen remain stable even after hydrogen diffuses out of the sample. Figure S8 hence depicts the dislocation cells observed on two different dates at the microscope within the span of approx. 2 months.

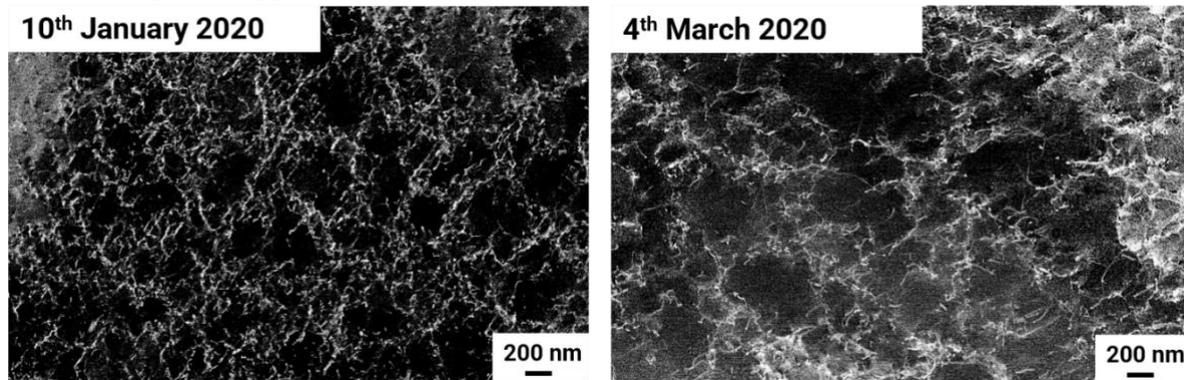

*Figure S8. The dislocation cells formed in the hydrogen-charged region of the 3% tensile strained sample observed first on 10th January 2020*